\documentstyle[12pt,epsfig]{article}
\makeatletter
\def\vereq#1#2{\lower3pt\vbox{\baselineskip1.5pt \lineskip1.5pt
\ialign{$\m@th#1\hfill##\hfil$\crcr#2\crcr\sim\crcr}}}
\makeatother
\begin{document}

\begin{titlepage}
\begin{flushright}
{DPNU-02-04}
\end{flushright}

\vskip 1cm
\begin{center}
{\large\bf Fermion Mass Hierarchy in 6 dimensional ${\mathbf{SO(10)}}$ SUSY GUT}
\vskip 1cm {\normalsize N. Haba$^{1,2}$, T. Kondo$^2$ and Y. Shimizu$^2$}
\\
\vskip 0.5cm {\it $^1$Faculty of Engineering, Mie University, Tsu, Mie,
  514-8507, Japan}\\
%\vskip 1cm {\normalsize T. Kondo and Y. Shimizu}\\
\vskip 0.5cm {\it $^2$Department of Physics, Nagoya University, Nagoya, 
  464-8602, Japan}
\end{center}
\vskip .5cm

%%%%%%%%%%%%%%%%%%%%%%%%%%%%%%%%%%%%%%%%%%%%%%
%%%%%%%%%%%%%%  ABSTRACT %%%%%%%%%%%%%%%%%%%%%
%%%%%%%%%%%%%%%%%%%%%%%%%%%%%%%%%%%%%%%%%%%%%%
\begin{abstract}

We suggest simple models which produce the suitable fermion
 mass hierarchies and flavor mixing angles based on the 
 supersymmetric $SO(10)$ grand unified theory 
 in 6 dimensions 
 compactified on a $T^2/Z_2$ orbifold.
The gauge and Higgs fields propagate in 6 dimensions while
 ordinal chiral matter fields are localized in 4 dimensions.
We introduce extra vector-like heavy fields in the extra 
 dimensions. 
We show three models according to the configurations 
 of vector-like fields in extra dimensions. 
The suitable fermion mass hierarchies and flavor mixings 
 are generated by integrating out
 these vector-like heavy fields.

\end{abstract}
\end{titlepage}
\setcounter{footnote}{0}
\setcounter{page}{1}
\setcounter{section}{0}
\setcounter{subsection}{0}
\setcounter{subsubsection}{0}

%%%%%%%%%%%%%%%%%%%%%%%%%%%%%%%%%%%%%%%%%%%%%%%%%%%%%%%%%%%%%%%%%%%
%%%%%%%%%%%%%%%%%%%%% INTRODUCTION %%%%%%%%%%%%%%%%%%%%%%%%%%%%%%%%
%%%%%%%%%%%%%%%%%%%%%%%%%%%%%%%%%%%%%%%%%%%%%%%%%%%%%%%%%%%%%%%%%%%
\section{Introduction}

\label{sec:intro}
Grand unified theories (GUTs) are very attractive models in which the three
gauge groups are unified at a high energy scale. 
However, one of the most serious problems to construct a model of GUTs 
is how to realize the mass splitting  between the triplet and 
 the doublet Higgs particles  in the Higgs sector. 
This problem is so-called  triplet-doublet (TD) splitting problem. 
Recently, a new idea for solving the TD splitting problem 
 has been suggested in 5 dimensional (5D) $SU(5)$ 
 GUT where the 5th dimensional 
 coordinate is compactified on an
 $S^1/(Z_2 \times Z_2')$ 
 orbifold \cite{5d}-\cite{ohio},
where only 
 Higgs and gauge fields can propagate in
 5 dimensions. 
The orbifolding makes the $SU(5)$ gauge group
 reduce to the SM gauge group and 
 realizes the TD splitting since  
 the doublet (triplet) Higgs fields have (not)
 Kaluza-Klein zero-modes. 
{}Following this new idea, 
a lot of works are made progress in the directions of 
 larger unified gauge symmetry and higher 
 space-time dimensions\cite{6d}-\cite{kawamoto}. 
In these models, gauge symmetry and supersymmetry are broken through 
 orbifold projections and/or the Scherk-Schwarz mechanism\cite{SS}.
Especially, the reduction of $SO(10)$ gauge symmetry 
and the TD splitting solution are 
 discussed in 6D models in Refs.\cite{ABC}\cite{HNOS}. 
%The gauge and Higgs fields 
%propagate in 6 dimensions, and the suitable
% orbifolding and boundary conditions on the fixed points 
% can induce the gauge symmetry
% reduction, $SO(10) \rightarrow 
% SU(3)_C\times SU(2)_L \times U(1)_Y \times U(1)_X$, 
% and realize the TD splitting by the similar mechanism as 
% the 5D $SU(5)$ GUT on $S^1/(Z_2 \times Z_2')$.

Several trials of producing fermion mass hierarchies 
 in the extra dimensional scenario have been 
 done in Refs.\cite{HSSU}\cite{HKS}\cite{flavor}. 
Especially, the model in Ref.\cite{HKS} can induce the natural
 fermion mass hierarchies and flavor mixings, 
 based on an $N = 1$ SUSY ((1,0)-SUSY) 
 $SO(10)$ GUT in 6 dimensions where
 the 5th and 6th dimensional coordinates are 
 compactified on a $T^2/Z_2$ orbifold\cite{HNOS}. 
The gauge and Higgs fields live in 6 dimensions 
while ordinal chiral matter fields are 
 localized in 4 dimensions\footnote{
$N = 1$ SUSY ((1,0)-SUSY) in 6 dimensions
requires the gauginos to have opposite chirality from 
the matter fermions which must all share the same chirality
\cite{others3}\cite{anomaly2}.
This feature strongly constrains our model 
when discussing 6D gauge anomaly. }. 
The orbifolding and boundary conditions make the $SO(10)$ 
 gauge group be broken to 
 $SU(3)_C\times SU(2)_L \times U(1)_Y \times U(1)_X$ and
 realize the TD splitting. 
In addition to the three-generation chiral matter
 fields, extra three sets of 
 vector-like matter fields have been introduced: 
 two sets of $\mathbf{16}$ and 
 $\overline{\mathbf{16}}$, and four $\mathbf{10}$ 
 representation fields\footnote{ 
 These $\mathbf{10}$ 
 representation matter fields should have 
 the same 6D chirality as $\mathbf{10}$ representation 
 Higgs fields for the 6D irreducible gauge anomaly 
cancellation\cite{others3}.}  
 propagate in 6 dimensions, and 
 chiral fields which couple to $\mathbf{16}$ and 
 $\overline{\mathbf{16}}$ 
 are defined as the 1st generation,
 and one set of $\mathbf{16}$ and 
 $\overline{\mathbf{16}}$ propagates 
 in 5 dimensions and chiral fields which couple to them are
 defined as the 2nd generation. 
%The introduction of the $\mathbf{16}$ and $\overline{\mathbf{16}}$ 
% Higgs multiplets on the 4D brane might break the $U(1)_X$ gauge 
% symmetry. 
%We assume the GUT scale of 
% $U(1)_X$ breaking, then suitable scale of 
% Majorana masses of right-handed 
% neutrinos are obtained.
The mixing angles between the chiral fields and 
 extra generations have been determined by
 the volume suppression factors. 
In this 
 paper, we suggest two more models as extensions of the 
 model in Ref.\cite{HKS} by introducing 
 additional vector-like matter 
 multiplets propagate in the bulk.
These models have different configurations of 
 additional vector-like fields in extra dimension. 
The suitable fermion mass hierarchies and flavor mixings 
 are generated by integrating out
 these vector-like heavy fields.

%%%%%%%%%%%%%%%%%%%%%%%%%%%%%%%%%%%%%%%%%%%%%%%%%%%%%%%%%%%%%%%%%%%
%%%%%%%%%%%%%            Section            %%%%%%%%%%%%%%%%%%%%%%%
%%%%%%%%%%%%%%%%%%%%%%%%%%%%%%%%%%%%%%%%%%%%%%%%%%%%%%%%%%%%%%%%%%%
\section{Framework in 6D ${\mathbf{SO(10)}}$ SUSY GUT}

We discussed the fermion mass hierarchy in the 6D $N=1$ SUSY 
((1,0)-SUSY) 
$SO(10)$ GUT \cite{HKS}.
The extra dimensions are compactified on a $T^2/Z_2$ orbifold, whose
coordinates ($x_5, x_6$) can be represented by the complex plane,
 $z \equiv x_5 + ix_6$. 
The structure of extra 2D spaces are 
characterized by reflection $P$ and translations $T_i$ ($i = 1, 2$). 
Under the reflection $P$, $z$ is transformed into $-z$, which corresponds
to the $\pi$ rotation on the complex plane. 
Under the translation  $T_i$, $z$ is transformed into 
$z + 2\pi R_5$ and $z + i 2\pi R_6$, respectively. 
$R_5$ and $R_6$ are the two radii of $T^2$.
We adopted the translations 
as $(T_1, T_2) = (T_{51}, T_{5'1'})$ where 
$T_{51} = \sigma_2\otimes I_5$ and $T_{5'1'} = 
\sigma_2\otimes diag.(1,1,1,-1,-1)$, which commute the generators
of the Georgi-Glashow $SU(5)\times U(1)_X$\cite{GG} and 
the flipped $SU(5)'\times U(1)'_X$\cite{Fl} groups, respectively
as in Refs.\cite{ABC}\cite{HNOS}.

When we construct a model on the 6D $N=1$ SUSY $SO(10)$ GUT,
we must pay attentions to following two characteristic features.
At first, 6D $N=1$ SUSY ((1,0)-SUSY) algebra 
determines the relation of 
6D chiralities between the
gauge, Higgs and matter multiplets, automatically.
6D $N = 1$ SUSY ((1,0)-SUSY) algebra requires 
the gauginos to have opposite chirality from 
the matter fermions which must all share the same 
 chirality\cite{others3}\cite{anomaly2}.
Thus, the 6D chiralities of all matter and Higgs 
fields which we introduced are the same each other, 
and opposite to those of gauginos.
Second, since the 6D theory is the chiral theory, 
 the 6D anomalies must be canceled 
by only the field propagating in 6 dimensions.
We know the 6D irreducible gauge anomaly do cancel between
a gauge multiplet and two ${\mathbf{10}}$ hypermultiplet or 
between a ${\mathbf{16}}$ (or  
${\overline{\mathbf{16}}}$) and a ${\mathbf{10}}$ hypermultiplet
\footnote{Other reducible anomalies can be canceled by the Green-Schwarz
mechanism \cite{GS}.}.

We consider the set-up that 
 the gauge multiplet and two ${\mathbf 10}$ representation 
 Higgs multiplets propagate in the 6D bulk and the ordinal 
 three-generation matter multiplets 
(${\mathbf{16}}_i, \;i=1,2,3$)
 localized on the 4D brane ($z = 0$). 
These field contents are free from gauge anomaly in 
6 dimensions\cite{others3}
%\cite{anomaly}
.
Moreover, the reflection(P) and translations($T_i$) make the $SO(10)$ 
 gauge group be broken to 
 $SU(3)_C\times SU(2)_L \times U(1)_Y \times U(1)_X$ and
 realize the TD splitting
 since the doublet (triplet) Higgs fields have (not)
 Kaluza-Klein zero-modes.

{}For the matter fields, 
 we introduced additional hypermultiplets,
 $\psi_{\bf 16}$ ($\psi_{\overline{{\bf 16}}}$), 
 in the bulk.
When the hypermultiplet $\psi_{\bf 16}$  propagates in the
6D bulk, we can classify $\psi_{\bf 16}$s into four types below,
\begin{eqnarray}
 {\psi_{\bf 16}}_{++}&~~~&\mbox{(zero mode)~}=Q, \nonumber\\
 {\psi_{\bf 16}}_{+-}&~~~&\mbox{(zero modes)}=\overline{U}
,\overline{E}, \nonumber\\
 {\psi_{\bf 16}}_{-+}&~~~&\mbox{(zero modes)}=\overline{D},
\overline{N}, \nonumber  \\
 {\psi_{\bf 16}}_{--}&~~~&\mbox{(zero mode)~}=L, 
\label{6d}
\end{eqnarray}
where the first $\pm$ sign represents the $T_1$ parity, while the
second $\pm$ sign represents the $T_2$ parity.
Equation (\ref{6d}) shows 
 zero-mode fields in each type of {\bf 16} 
 representation field. 
{}For the cancellation of the irreducible 6D gauge anomaly, 
 we must introduce  
 ${\mathbf 10}$ representation 
 hypermultiplets as many as the additional hypermultiplets,
 $\psi_{\bf 16}$ ($\psi_{\overline{{\bf 16}}}$). 
Similarly, when the hypermultiplet $\psi_{\bf 16}$  propagates in the
5D bulk ($x_6 = 0$),  we can classify $\psi_{\bf 16}$s into 
the two types below,
\begin{eqnarray}
 {\psi_{\bf 16}}_{+}&~~~&\mbox{(zero mode)~}=Q,\overline{U},
\overline{E}, \nonumber\\
 {\psi_{\bf 16}}_{-}&~~~&\mbox{(zero mode)}=L,\overline{D}
,\overline{N}, 
\label{5d}
\end{eqnarray}
where the $\pm$ sign represents the $T_1$ parity.

%%%%%%%%%%%%%%%%%%%%%%%%%%%%%%%%%%%%%%%%%%%%%%%%%%%%%%%%%%%%%%%%%%%%%
\subsection{Previous model}
\label{sec:MCAFMH}

Here let us review the previous model in Ref.\cite{HKS} 
 briefly. 
In the model we introduced ${\psi_{\bf 16}}_{++} + 
{\psi_{\overline{{\bf 16}}}}_{++} 
\equiv \psi_{{\bf 16}_{4}} + 
\psi_{\overline{{\bf 16}}_{4}}$ and ${\psi_{\bf 16}}_{+-} + 
{\psi_{\overline{{\bf 16}}}}_{+-}\equiv \psi'_{{\bf 16}_{4}} + 
\psi'_{\overline{{\bf 16}}_{4}}$ (we call them the 4th generation fields) 
 which propagate in the 6D bulk and 
interact with only the 1st generation matter multiplets.
{}For the 6D gauge anomaly cancellation, 
 four {\bf 10} representation matter fields 
 should also be introduced in 6D bulk, which 
 are heavy enough to have nothing to do with 
 the fermion mass hierarchies.  
In 5D bulk ($x_6 = 0$), we introduced 
 ${\psi_{\bf 16}}_{+} + {\psi_{\overline{{\bf 16}}}}_{+}
\equiv \psi_{{\bf 16}_{5}} + 
\psi_{\overline{{\bf 16}}_{5}}$, 
% which propagate in the 5D bulk and 
 (we call them the 5th generation fields) which 
 interact with only the 2nd generation matter multiplets.

\begin{table}
\begin{center}
\begin{tabular}{|c|c|} \hline
  matter multiplet     & additional multiplets  
\\ \hline
  ${\bf 16}_1$ & 
${\psi_{\bf 16}}_{++}$, ${\psi_{\bf 16}}_{+-}$  \\
  ${\bf 16}_2$ & 
${\psi_{\bf 16}}_{+}$    \\
  ${\bf 16}_3$ & \\
\hline
\end{tabular}
\end{center}
\caption{The matter fields in the previous model}
\label{additional matter} 
\end{table}

We put Higgs fields, ${H_{\bf 16}}$ and ${H_{\overline{\bf 16}}}$, 
 on 4D brane ($z=0$), which are assumed to take 
 vacuum expectation values (VEVs) of $O(10^{16})$ GeV 
 in the directions of 
 $B-L$. 
We impose Peccei-Quinn symmetry and its charge on the multiplets:
 all matter multiplets have its charge $1$, $\mathbf{10}$
 representation Higgs multiplets have its charge $-2$,
 and $\mathbf{16}$ and $\overline{\mathbf{16}}$ representation 
 Higgs multiplets have its charge $-1$.
The following interactions between the chiral 
 and extra generation fields on 4D brane, $z=0$, 
%$$
%W^{4D} \sim {1 \over M_*} {H_{\bf 16}}{H_{\overline{\bf 16}}}
%{\psi_{\bf 16}}{\psi_{\overline{\bf 16}}}
%$$
\begin{eqnarray}
\label{W6}
W_6 &=& H_{16}H_{\overline{16}} \left\{ 
\frac{y_{44}}{M_*^3}\psi_{{\bf 16}_4}\psi_{\overline{{\bf 16}}_4}
+ \frac{y'_{44}}{M_*^3}\psi'_{{\bf 16}_4}\psi'_{\overline{{\bf 16}}_4}
+ \frac{y_{14}}{M_*^{2}}{\bf 16}_1\psi_{\overline{{\bf 16}}_4}
+ \frac{y'_{14}}{M_*^{2}}{\bf 16}_1\psi'_{\overline{{\bf 16}}_4}
 \right.\nonumber\\
&& \left.
+ \frac{y_{55}}{M_*^2}\psi_{{\bf 16}_5}\psi_{\overline{{\bf 16}}_5}
+ \frac{y_{25}}{M_*^{3/2}}{\bf 16}_2\psi_{\overline{{\bf 16}}_5}\right\}
\delta(x_5)\delta(x_6),
\end{eqnarray}
induce the mass terms for the Kaluza-Klein zero-modes 
 of vector-like matter fields\footnote{
 As for {\bf 10} representation matter fields in 6D, 
 they have nothing to do with the fermion mass hierarchies 
 due to the PQ-symmetry and the mass terms are generated through 
 the interactions on $z=0$, 
$
W_{\psi_{\bf 10}} \sim {1 \over M_*^3} {H_{\bf 16}}{H_{\overline{\bf 16}}}
{\psi_{\bf 10}}{\psi_{\overline{\bf 10}}}\delta(x_5)\delta(x_6).
$}
as, 
\begin{eqnarray}
\label{W_4}
W_4 
&\simeq&
{v_N^2 \over M_*}
 \left\{\epsilon_1^{4}\left(Q^{(0)}_4 \overline{Q}^{(0)}_4
+ U'^{(0)}_4 \overline{U}'^{(0)}_4 + E'^{(0)}_4 \overline{E}'^{(0)}_4
\right) + \epsilon_1^{2} \left(Q_1 \overline{Q}^{(0)}_4  
+\overline{U_1} U'^{(0)}_4 + \overline{E_1} E'^{(0)}_4
\right)\right.  \nonumber\\
&&\left.
+ \epsilon_2^{2} \left(Q^{(0)}_5 \overline{Q}^{(0)}_5 
+ U^{(0)}_5 \overline{U}^{(0)}_5 + E^{(0)}_5 \overline{E}^{(0)}_5
\right) 
+ \epsilon_2 \left(Q_2 \overline{Q}^{(0)}_5 
+ \overline{U_2} U^{(0)}_5 + \overline{E_2} E^{(0)}_5
\right)
\right\},\label{W4}
\end{eqnarray}
where $\langle {H_{\bf 16}}\rangle =
 \langle{H_{\overline{\bf 16}}}\rangle \equiv v_N$ and 
$\epsilon_i$s are the volume suppression factors,
\begin{eqnarray} 
\epsilon_1 \equiv \sqrt{\frac{1}{2 \pi (R_5 R_6)^{1/2} M_*}}, \;\; \;\;  
\epsilon_2 \equiv \sqrt{\frac{1}{2 \pi R_5 M_*}}.
\label{eps}
\end{eqnarray}
Here we assume all Yukawa couplings are of order
 one. 
Thus the volume suppression factors $\epsilon_{1,2}$ 
 play crucial roles for generating 
 the fermion mass matrices 
 in the low energy. 
After integrating out 
 the heavy fields, 
 the model gives the following mass matrices 
 in the up quark sector, the down quark sector, 
 and the charged lepton sector\cite{HKS},
\begin{equation}
 m_u^l \simeq \left(
\begin{array}{ccc}
 \epsilon_1^4 & \epsilon_2 \epsilon_1^2 &  \epsilon_1^2  \\
 \epsilon_1^2 \epsilon_2 & \epsilon_2^2  &  \epsilon_2  \\
 \epsilon_1^2  & \epsilon_2  & 1
\end{array}
\right)  v, \;\;
 m_d^l \simeq \left(
\begin{array}{ccc}
 \epsilon_1^2 & \epsilon_1^2 &  \epsilon_1^2  \\
 \epsilon_2 & \epsilon_2  &  \epsilon_2  \\
  1 & 1 & 1 
\end{array}
\right) \overline{v}, \;\;
 m_e^l \simeq \left(
\begin{array}{ccc}
 \epsilon_1^2 & \epsilon_2 & 1  \\
 \epsilon_1^2 & \epsilon_2  & 1 \\
 \epsilon_1^2 & \epsilon_2 & 1 
\end{array}
\right)  \overline{v}, \;\;
\label{mass}
\end{equation}
respectively. 
 $\overline{v}$ and    
 $v$ are the vacuum expectation values of the weak Higgs doublets 
 in $\mathbf{10}$ representation Higgs multiplets.
We write the mass matrices in the basis
 that the left-handed fermions are to the left 
 and the right-handed fermions are to the right. 
We notice that all elements in the mass matrices 
 have $O(1)$ 
 coefficients.  
When we set $1/R_5 = 1/R_6 = O(10^{16})$ GeV, Eq.(\ref{eps}) gives 
$\epsilon_i \sim 0.04$ . 
Thus, we can regard $\epsilon_i \simeq\lambda^2$, 
where $\lambda$ is the Cabibbo angle, $\lambda \sim 0.2$. 
This model induces the fermion mass 
 hierarchies as, 
\begin{eqnarray}
\qquad\quad 
m_t : m_c : m_u  &\simeq & 1 : \lambda^4 : \lambda^8\;, \\  
m_b : m_s : m_d  &\simeq& m_{\tau} : m_{\mu} : m_e 
\simeq 1 : \lambda^2 : \lambda^4\;,
\end{eqnarray}
with the large $\tan \beta$. 
%We assume Yukawa couplings of Eq.(\ref{Yukawa1}) 
%to be $y \epsilon_{i}^2 = O(1)$. 
%Each element of Eq.(\ref{mass}) is understood to be 
%multiplied by $O(1)$ coefficient. 
%As a result the suitable mass hierarchies are 
%realized\cite{BB}-\cite{anarchy}.
%{}Moreover, 
% the small (large) flavor mixings 
% in the quark (lepton) sector are naturally 
% obtained. 
%
The mass matrix of three light neutrinos $m_\nu^{(l)}$ through
 the see-saw mechanism\cite{seesaw} as
\begin{equation}
\label{26}
 m_\nu^{(l)} \simeq
 {m_\nu^D m_\nu^D{}^T \over M_R} \simeq 
\left(
\begin{array}{ccc}
 1 & 1 & 1 \\
 1 & 1 & 1 \\
 1 & 1 & 1 
\end{array}
\right)  {v^2 \over M_R}.
\end{equation}
$M_R$ is about 
$10^{14}$ GeV induced from the interaction on $z=0$, 
\begin{equation}
\label{Yukawa2}
W_{M_N} = \frac{y^{N}_{ij}}{M_*} H_{\overline{16}}H_{\overline{16}} {\bf 
16}_i {\bf 16}_j \delta(x_5)\delta(x_6),
\end{equation}
%$$
%W^{4D} \sim {1 \over M_*} {H_{\overline{\bf 16}}}{H_{\overline{\bf 16}}}
%{\psi_{\bf 16}}_i{\psi_{\bf 16}}_j
%$$
where  $i,j=1 \sim 3$ are the generation indices.
We can obtain the suitable mass scale ($O(10^{-1})$ eV), for
 the atmospheric neutrino oscillation experiments, 
 taking account for the $SO(10)$ relation of 
 $y_u \simeq y_{\nu}$.

As for the flavor mixings, 
 the CKM\cite{CKM} and the MNS\cite{MNS} matrices are given by 
\begin{equation}
\label{mixing1}
 V_{CKM} \simeq \left(
\begin{array}{ccc}
 1 & \lambda^2 & \lambda^4 \\
 \lambda^2 & 1 & \lambda^2 \\
 \lambda^4 & \lambda^2 & 1 
\end{array}
\right)  , \;\;
 V_{MNS} \simeq \left(
\begin{array}{ccc}
 1 & 1 & 1 \\
 1 & 1 & 1 \\
 1 & 1 & 1 
\end{array}
\right) . 
\end{equation}
They give us a natural explanation 
 why the flavor mixing in the quark sector 
 is small while the flavor mixing in the 
 lepton sector is large\cite{BB}-\cite{anarchy}.
They suggest the suitable flavor mixings 
 roughly in order of magnitudes. 
However, they show too small Cabibbo angle and 
 too large $U_{e3}$. 
In order to obtain the suitable values of 
 them, we need suitable choice of $O(1)$ coefficients 
 in mass matrices.  
A suitable choice of $O(1)$ coefficients 
 in the mass matrix can really derive the suitable 
 flavor mixings consistent with the neutrino
 oscillation experiments as shown 
 in Ref.\cite{BB}. 
%Above fermion mass matrices yield the 
% suitable mass hierarchies of 
% quarks and leptons. 
%The above mass matrices are consistent with 
% experiments when we take account of $O(1)$ 
% coefficients. 
%The suitable 
% fermion mass hierarchies and mixing angles 
% both in the quark and the lepton sectors are really 
% induced from the above mass matrices with 
% the suitable $O(1)$ coefficients as in Ref.\cite{BB}. 
On the other hand, 
 if $O(1)$ coefficients are not determined by 
 a specific reason (symmetry)\footnote{
 This situation means the case when flavor mixing angles 
 are not very closed to maximals or zeros 
 but just large or small non-specific $O(1)$ values.  
 }, 
 it is meaningful to 
 see the most probable hierarchy and mixing angles 
 by considering random $O(1)$ coefficients\cite{anarchy}. 
However, if 
 the neutrino mixing angles are very closed 
 to be maximal or zero, 
 it is natural that 
 the mass matrices must be modified 
 to fit the experimental data 
 in {\it order} (power of $\lambda$) not by tunings of 
 $O(1)$ coefficients. 
We will discuss this trial in the following 
 two sections.

%%%%%%%%%%%%%%%%%%%%%%%%%%%%%%%%%%%%%%%%%%%%%%%%%%%%%%%%%%%%%%%%%%%
%%%%%%%%%%%%%            Section            %%%%%%%%%%%%%%%%%%%%%%%
%%%%%%%%%%%%%%%%%%%%%%%%%%%%%%%%%%%%%%%%%%%%%%%%%%%%%%%%%%%%%%%%%%%
\section{Model I}

{}For the extension of the model\cite{HKS}, 
we consider the situation where the 
 additional vector-like matter
 multiplets propagate in the bulk.
In model I, we introduced the matter multiplets, 
${\psi_{\bf 16}}_{-} + {\psi_{\overline{{\bf 16}}}}_{-}
\equiv \psi''_{{\bf 16}_{4}} + 
\psi''_{{\overline{\bf 16}}_{4}}$ 
(we call them the 4th generation fields)
which propagate in the 5D bulk ($x_6 = 0$) and 
interact with only the 1st generation matter multiplet
in addition to the previous model\cite{HKS}.
They have PQ charge $1$ as the other matter fields have. 

\begin{table}
\begin{center}
\begin{tabular}{|c|c|} \hline
  matter multiplet     & additional multiplets  
\\ \hline
  ${\bf 16}_1$ & 
${\psi_{\bf 16}}_{++}$, ${\psi_{\bf 16}}_{+-}$, 
${\psi_{\bf 16}}_{-}$  \\
  ${\bf 16}_2$ & 
${\psi_{\bf 16}}_{+}$    \\
  ${\bf 16}_3$ & \\
\hline
\end{tabular}
\end{center}
\caption{The matter fields in the Model I}
\label{additional matter1} 
\end{table}

In this case the following terms are added to Eq.(\ref{W6})
%{\bf please write Eq. corresponding to Eq.(3)}\\
\begin{eqnarray}
W_6 &=& H_{16}H_{\overline{16}} \left\{ 
 \frac{y''_{44}}{M_*^2}\psi''_{{\bf 16}_4}\psi''_{\overline{{\bf 16}}_4}
+ \frac{y''_{24}}{M_*^{3/2}}{\bf 16}_2\psi''_{\overline{{\bf 16}}_4}\right\}
\delta(x_5)\delta(x_6), 
\end{eqnarray}
then the term 
%{\bf please write Eq. corresponding to Eq.(4)}\\
\begin{eqnarray}
\label{W_42}
W_4 
&\simeq&
{v_N^2 \over M_*}
 \left\{
 \epsilon_2^{2} \left(L''^{(0)}_4 \overline{L}''^{(0)}_4 
+ D''^{(0)}_4 \overline{D}''^{(0)}_4 + N''^{(0)}_4 \overline{N}''^{(0)}_4
\right) 
 \right.\nonumber\\
&& \left.
+ \epsilon_2 \left(L_2 \overline{L}''^{(0)}_4 
+ \overline{D_2} D''^{(0)}_4 + \overline{N_2} N''^{(0)}_4
\right)
\right\},
\end{eqnarray}
is added to Eq.(\ref{W_4}). 
After integrating out the heavy vector-like fields, 
the mass matrices of the light eigenstates 
in the up quark sector, the down quark sector, 
and the charged lepton sector become     
\begin{equation}
 m_u^l \simeq \left(
\begin{array}{ccc}
 \epsilon_1^4 & \epsilon_2 \epsilon_1^2 &  \epsilon_1^2  \\
 \epsilon_1^2 \epsilon_2 & \epsilon_2^2  &  \epsilon_2  \\
 \epsilon_1^2  & \epsilon_2  & 1
\end{array}
\right)  v, \;\;
 m_d^l \simeq \left(
\begin{array}{ccc}
 \epsilon_1^2 \epsilon_2& \epsilon_1^2 &  \epsilon_1^2  \\
 \epsilon_2^2 & \epsilon_2  &  \epsilon_2  \\
  \epsilon_2 & 1 & 1 
\end{array}
\right) \overline{v}, \;\;
 m_e^l \simeq \left(
\begin{array}{ccc}
 \epsilon_1^2 \epsilon_2& \epsilon_2^2 & \epsilon_2  \\
 \epsilon_1^2 & \epsilon_2  & 1 \\
 \epsilon_1^2 & \epsilon_2 & 1 
\end{array}
\right)  \overline{v}, \;\;
\label{mass}
\end{equation}
respectively. 
The mass matrices  of the light eigenstates 
in the left-handed neutrino sector and the 
right-handed neutrino sector 
are given as   
\begin{equation}
 m_\nu^D \simeq \left(
\begin{array}{ccc}
\epsilon_2^2 & \epsilon_2 & \epsilon_2 \\
\epsilon_2 & 1 & 1 \\
\epsilon_2 & 1 & 1
\end{array}
\right)  v , \;\;
m_N \simeq \left(
\begin{array}{ccc}
\epsilon_2^2 & \epsilon_2 & \epsilon_2 \\
\epsilon_2 & 1 & 1 \\
\epsilon_2 & 1 & 1
\end{array}
\right) M_R .
\label{neu}
\end{equation}
Then 
the mass matrix of three light neutrinos $m_\nu^{(l)}$ through
the see-saw mechanism is given by 
\begin{equation}
\label{26}
 m_\nu^{(l)} \simeq
 {m_\nu^D m_\nu^D{}^T \over m_N} \simeq 
\left(
\begin{array}{ccc}
\epsilon_2^2 & \epsilon_2 & \epsilon_2 \\
\epsilon_2 & 1 & 1 \\
\epsilon_2 & 1 & 1
\end{array}
\right)  {v^2 \over M_R}.
\end{equation}
When we set $\epsilon_1 \sim \epsilon_2 \sim \lambda^2$, 
 which means $1/R_5 = 1/R_6 = O(10^{16})$ GeV, 
 we obtain the following fermion mass matrices\cite{lop} 
\begin{eqnarray}
 &m_u^l& \simeq \left(
\begin{array}{ccc}
 \lambda^8 & \lambda^6 &  \lambda^4  \\
 \lambda^6 & \lambda^4  &  \lambda^2  \\
 \lambda^4  & \lambda^2  & 1
\end{array}
\right)  v, \;\;
 m_d^l \simeq \left(
\begin{array}{ccc}
 \lambda^6 & \lambda^4 &  \lambda^4  \\
 \lambda^4 & \lambda^2  &  \lambda^2  \\
  \lambda^2 & 1 & 1 
\end{array}
\right) \overline{v}, \;\; \nonumber \\
 &m_e^l& \simeq \left(
\begin{array}{ccc}
 \lambda^6 & \lambda^4 & \lambda^2  \\
 \lambda^4 & \lambda^2  & 1 \\
 \lambda^4 & \lambda^2 & 1 
\end{array}
\right)  \overline{v}, \;\;
 m_\nu^{(l)} \simeq\left(
\begin{array}{ccc}
\lambda^4 & \lambda^2 & \lambda^2 \\
\lambda^2 & 1 & 1 \\
\lambda^2 & 1 & 1
\end{array}
\right)  {v^2 \over M_R}.
\end{eqnarray}
They induce the more suitable fermion 
 mass hierarchies in the power of $\lambda$ 
 for the down quark and 
 the charged lepton sectors, 
\begin{eqnarray}
\label{mass1}
\qquad\quad 
m_t : m_c : m_u  &\simeq & 1 : \lambda^4 : \lambda^8\;, \nonumber\\
m_b : m_s : m_d  &\simeq& m_{\tau} : m_{\mu} : 
m_e \simeq 1 : \lambda^2 : \lambda^6\;,
\end{eqnarray}
with large $\tan \beta$.
In order to obtain the realistic neutrino mass 
 hierarchy pattern which is consistent with neutrino
 oscillation experiments, 
 the rank of $2 \times 2$ sub-matrix in the 2nd and the 
 3rd generations in 
 $m_{\nu}^{(l)}$ should be reduced, and the 
 light eigenvalue of this sub-matrix should 
 be smaller than $O(\lambda^2)$. 
Then we might obtain the hierarchical type of 
 neutrino mass, $m_1 \ll m_2 \ll m_3$\cite{vissani}.

As for the flavor mixings, 
the CKM and the MNS matrices are given by 
\begin{equation}
 V_{CKM} \simeq \left(
\begin{array}{ccc}
 1 & \lambda^2 & \lambda^4 \\
 \lambda^2 & 1 & \lambda^2 \\
 \lambda^4 & \lambda^2 & 1 
\end{array}
\right) , \;\;
 V_{MNS} \simeq \left(
\begin{array}{ccc}
 1/\sqrt{2} & 1/\sqrt{2} & \lambda^2 \\
 1/2 & -1/2 & 1/\sqrt{2} \\
 -1/2 & 1/2 & 1/\sqrt{2} 
\end{array}
\right),
\end{equation}
where we assume the reduction of rank discussed above 
 in the neutrino 
 mass matrix, $m_\nu^{(l)}$, for obtaining this 
 MNS matrix. 
This type of flavor mixing matrix is so-called 
 bi-maximal one. 
Without the rank reduction, 
 the mixing between the 1st and the 2nd generations 
 can not be maximal. 
Since the value of $U_{e3}$ is of 
 order $\lambda^2$ in model I, 
 we do not need to choose $O(1)$ coefficients 
 as we did in the previous model. 
%the MNS matrix is just the same as
% that of the previous model in Eq.(\ref{mixing1}), and we should choose 
% the suitable parameter region of $O(1)$ coefficients 
% for the small value of $U_{e3}$. 
The CKM matrix has the same structure as in the previous model. 
This case shows 
 the relation between the Cabibbo angle 
 and quark masses, $\lambda^2 \sim \sqrt{m_d/m_s}$. 
Needless to say, the suitable choice of O(1) coefficients
 can reproduce the masses and the mixings 
 of quarks and leptons as in the previous model.

{}For another choice, 
 when we set 
% $\epsilon_1 \sim  \epsilon^{3/4} \sim \lambda^{3/2}$ 
 $\epsilon_1 \sim \lambda^{3/2}$ 
 and 
% $\epsilon_2 \sim \epsilon \sim \lambda^2$, 
 $\epsilon_2 \sim \lambda^2$, 
 which means $1/R_5=O(10^{16})$ GeV and $1/R_6=O(10^{17})$ GeV, 
 we obtain the following fermion mass matrices\cite{tani}, 
\begin{eqnarray}
 &m_u^l& \simeq \left(
\begin{array}{ccc}
 \lambda^6 & \lambda^5 & \lambda^3  \\
 \lambda^5 & \lambda^4 & \lambda^2  \\
 \lambda^3 & \lambda^2 & 1
\end{array}
\right)  v, \;\;
 m_d^l \;\simeq \left(
\begin{array}{ccc}
 \lambda^5 & \lambda^3 & \lambda^3  \\
 \lambda^4 & \lambda^2 & \lambda^2  \\
 \lambda^2 & 1 & 1 
\end{array}
\right) \overline{v}, \;\; \nonumber \\
 &m_e^l& \simeq \left(
\begin{array}{ccc}
 \lambda^5 & \lambda^4 & \lambda^2  \\
 \lambda^3 & \lambda^2 & 1 \\
 \lambda^3 & \lambda^2 & 1 
\end{array}
\right)  \overline{v}, \;\;
 m_\nu^{(l)} \simeq\left(
\begin{array}{ccc}
\lambda^4 & \lambda^2 & \lambda^2 \\
\lambda^2 & 1 & 1 \\
\lambda^2 & 1 & 1
\end{array}
\right)  {v^2 \over M_R}.
\end{eqnarray}
They suggest 
 the modified fermion mass hierarchies for the 
  quark and the charged lepton sectors, 
\begin{eqnarray}
\label{mass2}
\qquad\quad 
m_t : m_c : m_u  &\simeq & 1 : \lambda^4 : \lambda^6\; \nonumber\\ %= 
%1 : \epsilon^2 : \epsilon^3\;, \\
m_b : m_s : m_d  &\simeq& m_{\tau} : m_{\mu} : m_e \simeq 
1 : \lambda^2 : \lambda^5\; %=  
%1 : \epsilon : \epsilon^{5/2}\;,
\end{eqnarray}
with large $\tan \beta$.
The hierarchical neutrino mass pattern is obtained 
 when the rank of neutrino mass matrix is reduced 
 as in the previous case.

As for the flavor mixings, 
 the MNS matrix is the same as the 
 previous case. 
However, the CKM matrix is modified as, 
%As for the flavor mixings, 
%the CKM and the MNS matrices are given by 
\begin{equation}
 V_{CKM} \simeq \left(
\begin{array}{ccc}
 1 & \lambda & \lambda^3 \\
 \lambda & 1 & \lambda^2 \\
 \lambda^3 & \lambda^2 & 1 
\end{array}
\right). %\;\;
% V_{MNS} \simeq \left(
%\begin{array}{ccc}
% 1 & 1 & 1 \\
% 1 & 1 & 1 \\
% 1 & 1 & 1 
%\end{array}
%\right). 
\end{equation}
The hierarchal structure of the above CKM matrix 
nicely reproduces the experimental data\cite{PDG}.
The Cabibbo angle has the suitable value 
 with the 
 relation of $\lambda \sim \sqrt{m_d/m_s}$. 
Therefore we can conclude that the second choice
 of radius, $R_{5,6}$ is more suitable 
 for the fermion mass hierarchies and flavor
 mixings.

%%%%%%%%%%%%%%%%%%%%%%%%%%%%%%%%%%%%%%%%%%%%%%%%%%%%%%%%%%%%%%%%%%%
%%%%%%%%%%%%%            Section            %%%%%%%%%%%%%%%%%%%%%%%
%%%%%%%%%%%%%%%%%%%%%%%%%%%%%%%%%%%%%%%%%%%%%%%%%%%%%%%%%%%%%%%%%%%

\section{Model II}
As the second model, we introduce
 extra matter multiplets 
 in addition to the previous model\cite{HKS}: 
${\psi_{\bf 16}}_{-+} + {\psi_{\overline{{\bf 16}}}}_{-+} 
\equiv \psi'''_{{\bf 16}_{4}} + 
\psi_{{\overline{\bf 16}}'''_{4}}$ and 
${\psi_{\overline{{\bf 16}}}}_{--} + {\psi_{\overline{{\bf 16}}}}_{--}
 \equiv \psi''''_{{\bf 16}_{4}} + 
\psi''''_{{\overline{\bf 16}}_{4}}$ 
 (we call them the 4th generation fields) 
 which propagate in the 6D bulk\footnote{
 As commented in the previous discussion, 
 four additional {\bf 10} matter fields must be introduced for 
 the anomaly cancellation. However, we omit them since
 they have nothing to do with the fermion mass matrices
 due to the PQ-symmetry and their heavy masses.} and 
interact with only the 1st generation matter multiplet, 
${\psi_{\bf 16}}_{-} + {\psi_{\overline{{\bf 16}}}}_{-} 
\equiv \psi'_{{\bf 16}_{5}} + 
\psi'_{{\overline{\bf 16}}_{5}}$ 
(we call them the 5th generation fields)
which propagate in the 5D bulk ($x_6 = 0$) and 
interact with only the 2nd generation matter multiplet,
${\psi_{\bf 16}}_{-} + {\psi_{\overline{{\bf 16}}}}_{-}  
\equiv \psi_{{\bf 16}_{6}} + 
\psi_{{\overline{\bf 16}}_{6}}$
(we call them the 6th generation fields)
which propagate in the 5D bulk ($x_6 = 0$) and 
interact with only the 3rd generation matter multiplet.
All of them have $1$ PQ charges.

\begin{table}
\begin{center}
\begin{tabular}{|c|c|} \hline
  matter multiplet     & additional multiplets  
\\ \hline
  ${\bf 16}_1$ & 
${\psi_{\bf 16}}_{++}$, ${\psi_{\bf 16}}_{+-}$, 
${\psi_{\bf 16}}_{-+}$, ${\psi_{\bf 16}}_{--}$  \\
  ${\bf 16}_2$ & 
${\psi_{\bf 16}}_{+}$, ${\psi_{\bf 16}}_{-}$      \\
  ${\bf 16}_3$ & 
${\psi_{\bf 16}}_{-}$ \\
\hline
\end{tabular}
\end{center}
\caption{The matter fields in the Model II}
\label{additional matter2} 
\end{table}

In this case, the following terms are added to Eq.(\ref{W6}),
\begin{eqnarray}
W_6 &=& H_{16}H_{\overline{16}} \left\{ 
\frac{y'''_{44}}{M_*^3}\psi'''_{{\bf 16}_4}\psi'''_{\overline{{\bf 16}}_4}
+ \frac{y''''_{44}}{M_*^3}\psi''''_{{\bf 16}_4}\psi''''_{\overline{{\bf 16}}_4}
+ \frac{y'''_{14}}{M_*^{2}}{\bf 16}_1\psi'''_{\overline{{\bf 16}}_4}
+ \frac{y''''_{14}}{M_*^{2}}{\bf 16}_1\psi''''_{\overline{{\bf 16}}_4}
 \right.\nonumber\\
&+& \left.
 \frac{y'_{55}}{M_*^2}\psi'_{{\bf 16}_5}\psi'_{\overline{{\bf 16}}_5}
+ \frac{y'_{25}}{M_*^{3/2}}{\bf 16}_2\psi'_{\overline{{\bf 16}}_5}
+ \frac{y_{66}}{M_*^2}\psi_{{\bf 16}_6}\psi_{\overline{{\bf 16}}_6}
+ \frac{y_{36}}{M_*^{3/2}}{\bf 16}_3\psi_{\overline{{\bf 16}}_6}
\right\}\delta(x_5)\delta(x_6),~~~~~~ 
\end{eqnarray}
which induce additional terms to Eq.(\ref{W_4}) as,
\begin{eqnarray}
\label{W_43}
W_4 
&\simeq&
{v_N^2 \over M_*}
 \left\{\epsilon_1^{4}\left(D'''^{(0)}_4 \overline{D}'''^{(0)}_4
+ N'''^{(0)}_4 \overline{N}'''^{(0)}_4 
+ L''''^{(0)}_4 \overline{L}''''^{(0)}_4 \right) \right.  \nonumber\\
&+& \left.\epsilon_1^{2} \left(\overline{D}_1  D'''^{(0)}_4 
+\overline{N_1} N'''^{(0)}_4 +  L_1\overline{L_4}''''^{(0)}
\right)\right.  \nonumber\\
&+&
 \epsilon_2^{2} \left(D'^{(0)}_5 \overline{D}'^{(0)}_5 
+ N'^{(0)}_5 \overline{N}'^{(0)}_5 + L^{(0)}_5 \overline{L}'^{(0)}_5
\right) 
+ \epsilon_2 \left(\overline{D}_2 D'^{(0)}_5 
+ \overline{N_2} N'^{(0)}_5 + L_2 \overline{L}'^{(0)}_5
\right)
\nonumber\\
&+&\left.
 \epsilon_2^{2} \left(D^{(0)}_6 \overline{D}^{(0)}_6 
+ N^{(0)}_6 \overline{N}^{(0)}_6 + L^{(0)}_6 \overline{L}^{(0)}_6
\right) 
+ \epsilon_2 \left(\overline{D}_3 D'^{(0)}_6
+ \overline{N_3} N^{(0)}_6 + L_3 \overline{L}^{(0)}_6
\right)
\right\}.
\end{eqnarray}
After integrating out the vector-like heavy fields, 
the mass matrices of the light eigenstates 
in the up quark sector, the down quark sector, 
and the charged lepton sector are given as 
\begin{equation}
 m_u^l \simeq \left(
\begin{array}{ccc}
 \epsilon_1^4 & \epsilon_2 \epsilon_1^2 &  \epsilon_1^2  \\
 \epsilon_1^2 \epsilon_2 & \epsilon_2^2  &  \epsilon_2  \\
 \epsilon_1^2  & \epsilon_2  & 1
\end{array}
\right)  v, \;\;
 m_d^l \simeq \left(
\begin{array}{ccc}
 \epsilon_1^4 & \epsilon_1^2 \epsilon_2 &  \epsilon_1^2 \epsilon_2 \\
 \epsilon_2 \epsilon_1^2 & \epsilon_2^2  &  \epsilon_2^2  \\
  \epsilon_1^2 & \epsilon_2 & \epsilon_2 
\end{array}
\right) \overline{v}, \;\;
 m_e^l \simeq \left(
\begin{array}{ccc}
 \epsilon_1^4 & \epsilon_2\epsilon_1^2 & \epsilon_1^2  \\
 \epsilon_1^2 \epsilon_2 & \epsilon_2^2  & \epsilon_2 \\
 \epsilon_1^2 \epsilon_2 & \epsilon_2^2 & \epsilon_2 
\end{array}
\right)  \overline{v}, \;\;
\label{mass2}
\end{equation}
respectively. 
As the mass matrices  of the light eigenstates 
in the left-handed neutrino sector and the 
right-handed neutrino sector 
are given as   
\begin{equation}
 m_\nu^D \simeq \left(
\begin{array}{ccc}
\epsilon_1^4 & \epsilon_2\epsilon_1^2 & \epsilon_2\epsilon_1^2 \\
\epsilon_1^2\epsilon_2 & \epsilon_2^2 & \epsilon_2^2 \\
\epsilon_1^2\epsilon_2 & \epsilon_2^2 & \epsilon_2^2
\end{array}
\right)  v , \;\;
m_N \simeq \left(
\begin{array}{ccc}
\epsilon_1^4 & \epsilon_2\epsilon_1^2 & \epsilon_2\epsilon_1^2 \\
\epsilon_1^2\epsilon_2 & \epsilon_2^2 & \epsilon_2^2 \\
\epsilon_1^2\epsilon_2 & \epsilon_2^2 & \epsilon_2^2
\end{array}
\right)  M_R ,
\label{neu2}
\end{equation}
respectively, 
the mass matrix of three light neutrinos $m_\nu^{(l)}$ through
the see-saw mechanism is given by 
\begin{equation}
\label{light2}
 m_\nu^{(l)} \simeq
 {m_\nu^D m_\nu^D{}^T \over m_N} \simeq 
\left(
\begin{array}{ccc}
\epsilon_1^4 & \epsilon_2\epsilon_1^2 & \epsilon_2\epsilon_1^2 \\
\epsilon_1^2\epsilon_2 & \epsilon_2^2 & \epsilon_2^2 \\
\epsilon_1^2\epsilon_2 & \epsilon_2^2 & \epsilon_2^2
\end{array}
\right)  {v^2 \over M_R}.
\end{equation}
When we set $\epsilon_1 \sim \epsilon_2 \sim \lambda^2$, 
 which corresponds to $1/R_5 = 1/R_6 = O(10^{16})$ GeV, 
we obtain the following fermion mass matrices \cite{lop}
\begin{eqnarray}
 &m_u^l& \simeq \;\left(
\begin{array}{ccc}
 \lambda^8 & \lambda^6 &  \lambda^4  \\
 \lambda^6 & \lambda^4  &  \lambda^2  \\
 \lambda^4  & \lambda^2  & 1
\end{array}
\right)  v, \;\;
 m_d^l \;\;\simeq  \lambda^2 \;\left(
\begin{array}{ccc}
 \lambda^6 & \lambda^4 &  \lambda^4  \\
 \lambda^4 & \lambda^2  &  \lambda^2  \\
  \lambda^2 & 1 & 1 
\end{array}
\right) \overline{v}, \;\; \nonumber \\
 &m_e^l& \simeq  \lambda^2 \left(
\begin{array}{ccc}
 \lambda^6 & \lambda^4 & \lambda^2  \\
 \lambda^4 & \lambda^2  & 1 \\
 \lambda^4 & \lambda^2 & 1 
\end{array}
\;\right)  \overline{v}, \;\;
 m_\nu^{(l)} \simeq \lambda^4 \left(
\begin{array}{ccc}
\lambda^4 & \lambda^2 & \lambda^2 \\
\lambda^2 & 1 & 1 \\
\lambda^2 & 1 & 1
\end{array}
\;\right)  {v^2 \over M_R}.
\end{eqnarray}
The forms of these mass matrices are the same as 
 those of the first case of Model I except for 
 the overall factors. 
Thus the suitable fermion mass hierarchies 
 of the quark and the charged lepton sectors 
 are the same as Eq.(\ref{mass1}). 
The flavor mixing matrices, $V_{CKM}$ and $V_{MNS}$, 
 are also the same as those of the first case 
 of Model I. 
The different between this model and the first case of Model I 
 exists just in the value of 
 $\tan \beta$. 
This model shows the small $\tan \beta$ 
 of $\tan \beta \sim {m_t/m_b \over 1/\lambda^2} \sim 1$. 
The discussion of neutrino mass hierarchy 
 and the flavor mixings are also the same 
 as the first case of Model I.

{}For another choice, 
 when we set $\epsilon_1 \sim \lambda^{7/4}$ and 
 $\epsilon_2 \sim \lambda^2$,
 which means $1/R_5=O(10^{16})$ GeV and $1/R_6=O(10^{17})$ GeV, 
 we obtain the following fermion mass matrices.  
\begin{eqnarray}
 &m_u^l& \simeq \;\left(
\begin{array}{ccc}
 \lambda^7 & \lambda^{5 \mbox{-} 6} &  \lambda^{3 \mbox{-} 4}  \\
 \lambda^{5 \mbox{-} 6} & \lambda^4  &  \lambda^2  \\
 \lambda^{3 \mbox{-} 4}  & \lambda^2  & 1
\end{array}
\right)  v, \;\;
 m_d^l \;\;\simeq  \lambda^2 \;\left(
\begin{array}{ccc}
 \lambda^5 & \lambda^{3 \mbox{-} 4} &  \lambda^{3 \mbox{-} 4}  \\
 \lambda^{ 3\mbox{-} 4} & \lambda^2  &  \lambda^2  \\
  \lambda^{1 \mbox{-} 2} & 1 & 1 
\end{array}
\right) \overline{v}, \;\; \nonumber \\
 &m_e^l& \simeq  \lambda^2 \left(
\begin{array}{ccc}
 \lambda^5 & \lambda^{3 \mbox{-} 4} & \lambda^{1 \mbox{-} 2}  \\
 \lambda^{3 \mbox{-} 4} & \lambda^2  & 1 \\
 \lambda^{3 \mbox{-} 4} & \lambda^2 & 1 
\end{array}
\;\right)  \overline{v}, \;\;
 m_\nu^{(l)} \simeq \lambda^4 \left(
\begin{array}{ccc}
\lambda^{3} & \lambda^{1 \mbox{-} 2} & \lambda^{1 \mbox{-} 2} \\
\lambda^{1 \mbox{-} 2} & 1 & 1 \\
\lambda^{1 \mbox{-} 2} & 1 & 1
\end{array}
\;\right)  {v^2 \over M_R}.
\end{eqnarray}
This case also shows 
 the small $\tan \beta$. 
These mass matrices induce the fermion mass hierarchies 
 of the quark and the charged lepton sectors 
 as 
\begin{eqnarray}
\label{mass11}
\qquad\quad 
m_t : m_c : m_u  &\simeq & 1 : \lambda^4 : \lambda^7\;, \nonumber\\
m_b : m_s : m_d  &\simeq& m_{\tau} : m_{\mu} : 
m_e \simeq 1 : \lambda^2 : \lambda^5\;. 
\end{eqnarray}
The flavor mixing matrices are given by 
\begin{equation}
 V_{CKM} \simeq \left(
\begin{array}{ccc}
 1 & \lambda^{1 \mbox{-} 2} & \lambda^{3 \mbox{-} 4} \\
 \lambda^{1 \mbox{-} 2}& 1 & \lambda^2 \\
 \lambda^{3 \mbox{-} 4} & \lambda^2 & 1 
\end{array}
\right) , \;\;
 V_{MNS} \simeq \left(
\begin{array}{ccc}
 1/\sqrt{2} & 1/\sqrt{2} & \lambda^2 \\
 1/2 & -1/2 & 1/\sqrt{2} \\
 -1/2 & 1/2 & 1/\sqrt{2} 
\end{array}
\right).
\end{equation}
Here we assume the reduction of rank in the neutrino 
mass matrix, $m_\nu^{(l)}$, for obtaining this 
MNS matrix.

%%%%%%%%%%%%%%%%%%%%%%%%%%%%%%%%%%%%%%%%%%%%%%%%%%%%%%%%%%%%%%%%%
%%%%%%%%%%%%  Summary     %%%%%%%%%%%%%%%%%%%%%%%%%%%%%%%%%%%%%%%
%%%%%%%%%%%%%%%%%%%%%%%%%%%%%%%%%%%%%%%%%%%%%%%%%%%%%%%%%%%%%%%%% 
\section{Summary and Discussion}

In this 
 paper, we have shown three models based on 
 the 6D $N = 1$ SUSY 
 $SO(10)$ GUT where the 5th and 6th dimensional coordinates are 
 compactified on a $T^2/Z_2$ orbifold. 
The gauge and Higgs fields live 
 in 6 dimensions while ordinal chiral matter fields are 
 localized in 4 dimensions. 
%The orbifolding and boundary conditions make the $SO(10)$ 
% gauge group be broken to 
% $SU(3)_C\times SU(2)_L \times U(1)_Y \times U(1)_X$, and
% realize the TD splitting. 
At first, we have shown briefly the model in
 Ref.\cite{HKS}
 where in addition to the three-generation chiral matter
 fields, extra three sets of vector-like  
 matter fields are introduced: 
 $\psi_{{\bf 16}_4}+ \psi_{\overline{{\bf 16}}_4}$, 
 $\psi'_{{\bf 16}_4}+ \psi'_{\overline{{\bf 16}}_4}$ 
 and four $\psi_{\mathbf{10}_{\alpha}}$s 
 propagate in 6 dimensions and 
 chiral fields which couple to 
 $\psi_{{\bf 16}_4}+ \psi_{\overline{{\bf 16}}_4}$, 
 $\psi'_{{\bf 16}_4}+ \psi'_{\overline{{\bf 16}}_4}$ 
 are defined as the 1st generation, and
 $\psi_{{\bf 16}_5}+ \psi_{\overline{{\bf 16}}_5}$ propagate 
 in 5 dimensions and chiral fields which couple to them are
 defined as the 2nd generation.
This model shows the fermion mass hierarchies as, 
$m_t : m_c : m_u  \simeq  1 : \lambda^4 : \lambda^8\;, 
m_b : m_s : m_d  \simeq m_{\tau} : m_{\mu} : m_e 
\simeq 1 : \lambda^2 : \lambda^4\;.
$
and flavor mixings as, 
%\begin{equation}
$$
 V_{CKM} \simeq \left(
\begin{array}{ccc}
 1 & \lambda^2 & \lambda^4 \\
 \lambda^2 & 1 & \lambda^2 \\
 \lambda^4 & \lambda^2 & 1 
\end{array}
\right)  , \;\;
 V_{MNS} \simeq \left(
\begin{array}{ccc}
 1 & 1 & 1 \\
 1 & 1 & 1 \\
 1 & 1 & 1 
\end{array}
\right) . 
%\end{equation}
$$

We suggest two more models as the extensions 
of this model. 
The model I has additional  
 ${\psi_{\bf 16}}_{-} + {\psi_{\overline{{\bf 16}}}}_{-}$ 
 matter fields propagating in 5D space, $x_6=0$, 
 which couple to the 1st generation chiral
 fields.
When we set $\epsilon_1 \sim \epsilon_2 \sim \lambda^2$,
 which means $1/R_5 = 1/R_6 = O(10^{16})$ GeV, 
this model shows the mass hierarchies in the quark 
 and the charged lepton sectors as, 
$m_t : m_c : m_u  \simeq  1 : \lambda^4 : \lambda^8\;, 
m_b : m_s : m_d  \simeq m_{\tau} : m_{\mu} : m_e 
\simeq 1 : \lambda^2 : \lambda^6\;.$
When we assume that the rank is reduced in the 
 neutrino mass matrix, the mass spectra 
 of the neutrino mass eigenvalues 
 becomes hierarchical, and the 
 flavor mixings are given by 
%\begin{equation}
$$
 V_{CKM} \simeq \left(
\begin{array}{ccc}
 1 & \lambda^2 & \lambda^4 \\
 \lambda^2 & 1 & \lambda^2 \\
 \lambda^4 & \lambda^2 & 1 
\end{array}
\right) , \;\;
 V_{MNS} \simeq \left(
\begin{array}{ccc}
 1/\sqrt{2} & 1/\sqrt{2} & \lambda^2 \\
 1/2 & -1/2 & 1/\sqrt{2} \\
 -1/2 & 1/2 & 1/\sqrt{2} 
\end{array}
\right). 
%\end{equation}
$$
{}For another choice of the sizes of compactification length 
 of the extra dimensions as 
 $\epsilon_1 \sim \lambda^{3/2}$ 
 and $\epsilon_2 \sim \lambda^2$, which means 
 $1/R_5=O(10^{16})$ GeV and $1/R_6=O(10^{17})$ GeV, 
 this case shows the mass hierarchies of 
 the quark and the charged lepton sectors as, 
%{\bf Plaease write here!}   
$m_t : m_c : m_u  \simeq  1 : \lambda^4 : \lambda^6\;, 
m_b : m_s : m_d  \simeq m_{\tau} : m_{\mu} : m_e 
\simeq 1 : \lambda^2 : \lambda^5\;.$
The case with the rank reduction of the neutrino 
 mass matrix can give the hierarchical neutrino 
 masses and the 
 flavor mixings as, 
%\begin{equation}
$$
 V_{CKM} \simeq \left(
\begin{array}{ccc}
 1 & \lambda & \lambda^3 \\
 \lambda & 1 & \lambda^2 \\
 \lambda^3 & \lambda^2 & 1 
\end{array}
\right), \;\;
 V_{MNS} \simeq \left(
\begin{array}{ccc}
 1/\sqrt{2} & 1/\sqrt{2} & \lambda^2 \\
 1/2 & -1/2 & 1/\sqrt{2} \\
 -1/2 & 1/2 & 1/\sqrt{2} 
\end{array}
\right). 
%\end{equation}
$$
The too small (large) value of the
Cabibbo angle ($U_{e3}$) can be avoided
by power of $\lambda$.

The model II has additional  
${\psi_{\bf 16}}_{++} + 
{\psi_{\overline{{\bf 16}}}}_{++}$ and ${\psi_{\bf 16}}_{+-} + 
{\psi_{\overline{{\bf 16}}}}_{+-}$
 which propagate in the 6D bulk and 
interact with only the 1st generation matter multiplets, and 
 ${\psi_{\bf 16}}_{+}$ (${\psi_{\overline{{\bf 16}}}}_{+}$), 
% which propagate in the 5D bulk and 
  in 5D bulk ($x_6 = 0$) 
  which interact with only the 2nd generation matter multiplets. 
When we set $\epsilon_1 \sim \epsilon_2 \sim \lambda^2$,
 which means $1/R_5 = 1/R_6 = O(10^{16})$ GeV, 
 the fermion mass hierarchies 
 and flavor mixings are the same as 
 the first case of Model I. 
This case, however, suggests small $\tan \beta$ 
 contrary to the first case of
 Model I. 
{}For another choice of 
 $\epsilon_1 \sim \lambda^{7/4}$ and 
 $\epsilon_2 \sim \lambda^2$,
 which means $1/R_5=O(10^{16})$ GeV and $1/R_6=O(10^{17})$ GeV, 
 the fermion mass hierarchies are given by 
$m_t : m_c : m_u  \simeq  1 : \lambda^4 : \lambda^7, \; 
m_b : m_s : m_d  \simeq m_{\tau} : m_{\mu} : 
m_e \simeq 1 : \lambda^2 : \lambda^5\; $ with 
 small $\tan \beta$. 
Assuming the rank reduction in the 
 neutrino mass matrix, the flavor mixing 
 matrices are given by 
%
%\begin{equation}
$$
 V_{CKM} \simeq \left(
\begin{array}{ccc}
 1 & \lambda^{1 \mbox{-} 2} & \lambda^{3 \mbox{-} 4} \\
 \lambda^{1 \mbox{-} 2}& 1 & \lambda^2 \\
 \lambda^{3 \mbox{-} 4} & \lambda^2 & 1 
\end{array}
\right) , \;\;
 V_{MNS} \simeq \left(
\begin{array}{ccc}
 1/\sqrt{2} & 1/\sqrt{2} & \lambda^2 \\
 1/2 & -1/2 & 1/\sqrt{2} \\
 -1/2 & 1/2 & 1/\sqrt{2} 
\end{array}
\right). 
$$
%\end{equation}
This case can also avoid 
 too small value of the
 Cabibbo angle.

%%%%%%%%%%%%%%%%%%%%%%%%%%%%%%%%%%%%%%%%%%%%%%%%%%%%%%%%%%%%%%%%%%%
%%%%%%%%%%%%%       Acknowledgment          %%%%%%%%%%%%%%%%%%%%%%%
%%%%%%%%%%%%%%%%%%%%%%%%%%%%%%%%%%%%%%%%%%%%%%%%%%%%%%%%%%%%%%%%%%%

\section*{Acknowledgment}
We would like to thank M. Bando and 
 N. Maekawa for useful discussions at 
 the workshop ``post NOON" in YITP on 
 December 2001. 
This work is supported in
 part by the Grant-in-Aid for Science
 Research, Ministry of Education, Culture, Sports, Science and 
Technology, of Japan (No.12740146, No.13001292).

%%%%%%%%%%%%%%%%%%%%%%%%%%%%%%%%%%%%%%%%%%%%%%%%%%%%%%%%%%%%%%%%%
%%%%%%%%%%%%  References  %%%%%%%%%%%%%%%%%%%%%%%%%%%%%%%%%%%%%%%
%%%%%%%%%%%%%%%%%%%%%%%%%%%%%%%%%%%%%%%%%%%%%%%%%%%%%%%%%%%%%%%%%

\end{document}